\documentclass[prl,reprint,superscriptaddress,amssymb,amsmath,amsfonts,nofootinbib,showpacs]{revtex4-1}
\pdfoutput=1 
\usepackage{graphicx,epsf, epsfig, amssymb}
\usepackage{color}
\usepackage{times}
\usepackage{mathptmx}
\usepackage{hyperref}
\usepackage{mathrsfs}
\usepackage[normalem]{ulem}

\def\be{\begin{equation}}
\def\ee{\end{equation}}
\def\beq{\begin{eqnarray}}
\def\eeq{\end{eqnarray}}

\newcommand{\PhiRet}{\Phi^{\rm ret}}
\newcommand{\PhiR}{\Phi^{\rm R}}
\newcommand{\PhiS}{\Phi^{\rm S}}
\newcommand{\scri}{\mathscr{I}}

\begin{document}

\title{Self-consistent orbital evolution of a particle around a Schwarzschild black hole}


\author{Peter Diener}
\affiliation{Center for Computation \& Technology, Louisiana State University, Baton Rouge, LA 70803, U.S.A.}

\affiliation{Department of Physics \& Astronomy, Louisiana State University, Baton Rouge, LA 70803, U.S.A.}

\author{Ian Vega}
\affiliation{Department of Physics, University of Guelph, Guelph, Ontario, N1G 2W1, Canada}

\author{Barry Wardell}
\affiliation{Max-Planck-Institut f\"{u}r Gravitationphysik, Albert-Einstein-Institut, 14476 Potsdam, Germany}
\affiliation{School of Mathematical Sciences and Complex \& Adaptive Systems Laboratory,\\ University College Dublin, Belfield, Dublin 4 Ireland}

\author{Steven Detweiler}
\affiliation{Institute for Fundamental Theory, Department of Physics, University of Florida, Gainesville, FL 32611-8440, U.S.A.}

\begin{abstract}
The motion of a charged particle is influenced by the self-force arising from the particle's interaction
with its own field.
In a curved spacetime, this self-force depends on the entire 
past history of the particle and is difficult to evaluate. As a 
result, all existing self-force evaluations in curved spacetime 
are for particles moving along a fixed trajectory.
Here, for the first time, we overcome this long-standing limitation and present fully 
self-consistent orbits and waveforms of a scalar charged particle around a Schwarzschild black hole. 
\end{abstract} 

\maketitle
In spite of the impressive progress made towards tackling the two-body 
problem in general relativity \cite{pretorius:05,*campanelli-etal:06,*baker-etal:06,*lousto-zlochower:11}, 
there remains an important regime that appears to be intractable by the methods of numerical relativity. When 
the system consists of a massive black hole ($M \gtrsim 10^6 
M_{\odot}$) and a stellar mass companion ($m \sim 10M_{\odot}$), the 
disparity of length scales characterizing this black hole binary proves
to be a significant roadblock for existing numerical relativity codes.

The dynamics of such a binary is intuitively simple:
A slow adiabatic inspiral of the small black hole towards the bigger one 
is followed by an abrupt plunge towards the latter's 
event horizon. However, for the purposes of gravitational wave astronomy, 
this qualitative picture is inadequate. The ubiquity of supermassive 
black holes  residing in galactic centers has made 
extreme-mass-ratio inspirals (EMRIs) one 
of the more prominent predicted sources of low-frequency
gravitational waves for future space-based missions \cite{amaro-seoane-etal:07}.
The science we will gain from these sources --- among them precision tests 
of general relativity in the strong-field regime \cite{ryan:95,*collins-hughes:04,*glampedakis-babak:06,*gair-etal:08} 
and a better census of black hole populations \cite{barack-cutler:04,amaro-seoane-etal:07} --- rests
on our ability to model them to an exquisite degree of accuracy. 
Specifically, we wish to be able to track the phase of their gravitational 
waveforms throughout the long inspiral. 


In the self-force approach to modeling EMRIs, one ignores the 
internal dynamics of the smaller black hole and treats it as a 
massive particle that distorts the spacetime geometry of the bigger 
partner. An EMRI is then equivalent to a charged 
particle moving in a black hole spacetime.
But for this approach to suffice, the motion of the particle 
and the resulting waveform need to incorporate self-force effects 
arising from the interaction of the particle with its own field. 

Evaluating the self-force is a 
difficult, though by now well understood, process \cite{LLR:11,barack:09}. 
In a curved spacetime, the field 
generated by a particle at one time 
backscatters off the curvature and interacts with
the particle at a much later time. Consequently, the
self-force at any given instant depends on the particle's 
entire past history \cite{dewitt-brehme:60,*mino-etal:97,*quinn-wald:97}. 
This restricts the usefulness of purely analytical self-force 
calculations mainly to astrophysically uninteresting 
cases \cite{smith-will:80,*wiseman:00}. 
On the other hand, the distributional nature of the point 
source makes numerical evaluation of the self-force 
technically involved. The retarded field diverges at the location 
of the particle, thus requiring a delicate regularization  
to extract the finite self-force \cite{barack:09}.
A practical scheme for dealing with this difficulty exists; 
this is the spherical-harmonic-based mode-sum method of 
Barack and Ori \cite{barack-ori:00}. 
This method has been tremendously successful for self-force calculations based on a 
specified particle orbit \cite{barack:00,*barack-burko:00,*barack-lousto:02,*burko:00c,
*diazrivera-etal:04,*barack-sago:07,*detweiler:08,*warburton-barack:10, 
*warburton-barack:11,haas-poisson:06,*detweiler-etal:03,haas:07,barack-sago:10}. However, 
it has not yet been applied to compute the self-force based on an evolved orbit.


A problem that has resisted solution for a long time is 
the computation of \emph{self-consistently} 
self-forced orbits and their corresponding waveforms. These are orbits that reflect the true 
motion of the particle as it is driven by its actual local field. 
In principle, these self-consistent orbits can be obtained only by 
simultaneously solving the equations governing the coupled dynamics of the 
particle and its field. 

This notion of self-consistency is what we wish to highlight, 
particularly because a recent manuscript \cite{warburton-etal:12} 
also reports on self-forced orbits, though not of the sort we present 
here.
In that work, the applied self-force at some instant 
is not what arises from the actual field at that same instant. 
Instead, this applied self-force is what would have resulted if the 
particle were moving for all eternity along the geodesic 
that only instantaneously matches the true orbit; it is 
the ``geodesic" self-force and not the self-consistent self-force.
It is quite likely that the error incurred by this assumption 
becomes negligible in the adiabatic 
limit, for which the particle stays close 
to the instantaneous geodesic for sufficiently long times. 
We emphasize, however, that this is presently just 
an expectation rather than a demonstrated fact.
Rigorously assessing its validity requires 
comparison with fully self-consistent orbits.

\begin{figure}[t!]
\begin{center}
\includegraphics[clip,width=8cm,angle=0]{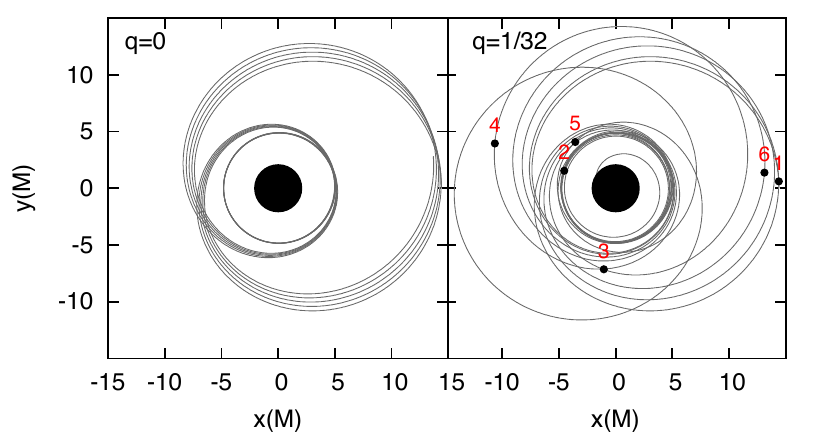}
\caption{(Color online) Orbits for neutral and charged
particles starting at $p=7.2$, $e=0.5$. The orbital evolution is started close
to apastron (at $t=t_1$) and the dots represent events at times
$t_1=400M, t_2 = 600M, t_3= 1100M, t_4 = 1300M, t_5 = 1800M$ and $t_6 = 2043.8M$, the instant of plunge..
The coordinates $\{x,y\}=\{r\cos\phi,r\sin\phi\}$ are Cartesian coordinates in the equatorial plane.}
\label{fig:orbit}
\end{center}
\end{figure}

\begin{figure}[t]
\begin{center}
\includegraphics[clip,width=8.5cm,angle=0]{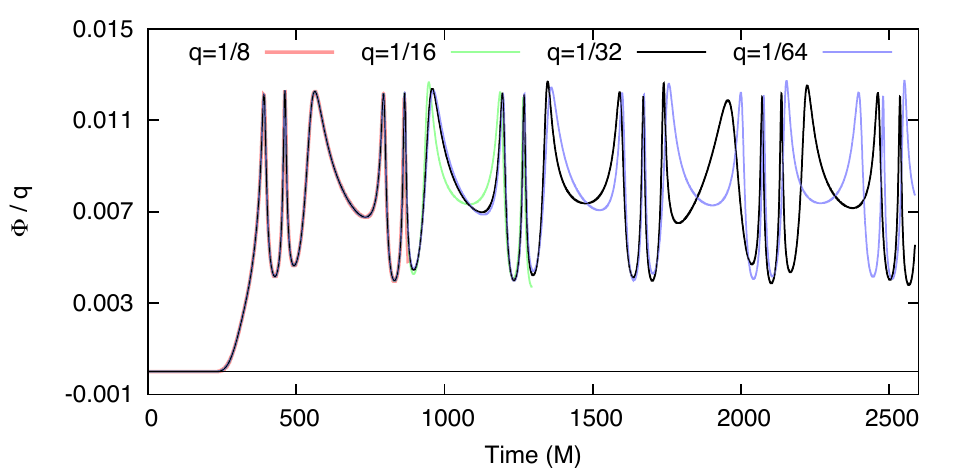}
\caption{(Color online) Waveforms from self-consistently evolved orbits as detected 
by an observer located in the orbital plane at ${\scri}^{+}$.}
\label{fig:fluxes}
\end{center}
\end{figure}

In this \emph{Letter}, we present for the first time such fully 
self-consistent orbits and waveforms, albeit for a radiating scalar charge 
in the Schwarzschild spacetime. An example 
is displayed in Figs. \ref{fig:orbit} and \ref{fig:fluxes}.

\emph{Effective source approach}:-- A novel strategy for
self-force calculations was proposed
\cite{vega-detweiler:08,*vega-etal:09,*vega-etal:11} to
address the difficulties 
arising from a $\delta$-function source. Its core
idea is to refrain from solving the retarded field
altogether and to work instead with the equation for a 
field $\PhiR$ from which the self-force
is readily computed: $f_\alpha \equiv \bar{q}\nabla_\alpha \PhiR$, where
$\bar{q}$ is the charge of the particle. The \emph{effective} source $S(x^\alpha|z^\alpha(\tau),u^\alpha(\tau))$
for $\PhiR$ is $C^0$ (continuous but not differentiable) by construction 
at the location of the
particle, in contrast to the traditional $\delta$-function
source for point particles. Like the $\delta$-function source, $S$
depends on the particle's position $z^\alpha(\tau)$ and
four-velocity $u^\alpha(\tau)$. (A
similar approach is independently being pursued in
Ref.~\cite{barack-golbourn:07,*barack-etal:07,*dolan-barack:11,*dolan-etal:11},
the main difference there being that it
uses a mode decomposition in the azimuthal direction).
The strategy rests on Detweiler and Whiting's insight \cite{detweiler-whiting:03}
that there exists a smooth solution to the vacuum field equation to
which the self-force can be fully attributed. This solution is just
the difference between the retarded field $\PhiRet$ and a
locally constructible singular field $\PhiS$ which is the curved spacetime
analogue of a ``Coulomb'' field that does not contribute to the self-force.
Our approximation to the regular field $\PhiR$ differs from the smooth
Detweiler-Whiting solution by terms that scale as $O(\rho^3)$ as
$\rho \rightarrow 0$, where $\rho$ is some appropriate
measure of distance from the particle. It is thus only a
$C^2$ approximation to the Detweiler-Whiting vacuum solution, but it nevertheless
gives the same self-force. The limited differentiability 
of $\PhiR$ comes from the inability to write down an explicit expression for the full singular field 
from which the effective source is constructed. Generally, only approximate expressions for
$\PhiS$ are known\cite{haas-poisson:06,*detweiler-etal:03}.
The construction of our 
expression for $S$ is described in detail
elsewhere
\cite{wardell-etal:12}.

With the effective source at hand, one needs to solve the 
following system of equations:
\begin{align}
    \Box\PhiR &= S(x|z(\tau),u(\tau)) \label{eqn:fieldEqn}\\
    \frac{Du^\alpha}{d\tau} &= a^\alpha =
        \frac{\bar{q}}{m(\tau)}(g^{\alpha\beta}+u^\alpha u^\beta)\nabla_\beta
                \PhiR \label{eqn:EOM}\\
    \frac{dm}{d\tau} &= 
                -\bar{q}u^\beta\nabla_\beta
                \PhiR, \label{eqn:massEqn} 
\end{align} 
where $m(\tau)$ is the rest mass of the particle. Quinn \cite{quinn:00} 
found that the rest mass is dynamically modified by the component of the self-force 
tangent to the four-velocity; 
this is reflected in Eq. (\ref{eqn:massEqn}). 
In all our simulations, we take the initial rest mass $m(\tau=0)$ to be $M$.  
Because of the way $S$ is constructed,
$\PhiR$ is equal to the retarded field in the wavezone. Thus, by 
solving the system of equations above, one obtains not only self-forced 
orbits but their corresponding waveforms as well
(see Figs. \ref{fig:orbit} and \ref{fig:fluxes}).

We recently developed code that 
solves Eq. (\ref{eqn:fieldEqn}) for a specified 
geodesic in the Schwarzschild spacetime \cite{diener-etal:12b}.
Comparing with Ref.~\cite{haas:07} we find that our main source of error is high
frequency noise due to nonsmoothness of the effective source in the vicinity
of the worldline. Most of the time the amplitude of this noise is small but it
reaches a peak of about 2\% of the value of the self-force at periapsis.

We then 
evolve self-consistent orbits by supplementing 
the scalar field evolution with an orbit integrator. Together 
they allow solving Eqs. (\ref{eqn:fieldEqn}),
(\ref{eqn:EOM}) and (\ref{eqn:massEqn}) simultaneously.  
We deal with the particle motion in two ways: first 
by straightforward
integration of Eq.~(\ref{eqn:EOM}) and second by adopting the osculating orbits framework described 
in Ref.~\cite{pound-poisson:08b}. The first method is more general in that
it allows us to track the motion of the particle all the way to the 
event horizon. On the other hand, the second method, which works only 
for bound orbits, allows us to more readily 
identify aspects of the evolution that would be completely missed by 
methods relying on flux-averaging and balance arguments. 
For the regimes in which both methods are valid, we find the resulting
orbits in excellent agreement.

\emph{Self-consistent orbits}:-- 
The spherical symmetry of the Schwarzschild geometry implies that 
test-particle orbits may always be described by motion 
in the $\theta = \pi/2$ plane. 
The orbits are then characterized by conserved quantities $\tilde{E}$ 
and $\tilde{L}$, the particle's energy and angular momentum per unit mass, respectively. 
Bound orbits are those for which 
$\tilde{E} < 1$ and $\tilde{L} \geq 2\sqrt{3}M$. These 
possess two radial turning points $r_\pm$ 
($r_- < r_+$), the periapsis and apoapsis. Following 
Refs.~\cite{pound-poisson:08b,cutler-etal:94}, 
these orbits can be parametrized 
in terms of a dimensionless semilatus rectum $p$ and 
eccentricity $e$, such that $r_\pm = pM/(1\mp e)$.
This $p$--$e$ parametrization is geometrically informative: 
$p$ is a measure of the size of the orbit, while $e$ 
is a measure of deviation from circularity. We note,
however, that it is meaningful only for the space 
of bound orbits for which
$\{\tilde{E} < 1,\tilde{L} >2\sqrt{3}M\}$
is mapped onto $\{0\leq e<1, p \geq 6 + 2e\}$.
The separatrix $p=6+2e$ corresponds to unstable circular orbits and 
represents the boundary in $p$--$e$ space separating 
bound from plunging orbits.

In this parametrization, orbits are described by
\begin{equation}
    r(t) = \frac{Mp}{1+e\cos(\chi-w)}
\end{equation}
\begin{align}
\frac{d\phi}{dt} = \bigg[1&-\frac{2M r'}{r-2M}\bigg] \times \nonumber \\
  & \frac{[p-2-2e \cos(\chi-w)][1+e\cos(\chi-w)]^2}{M \sqrt{p^3[(p-2)^2-4e^2]}},
\end{align}
where $r \equiv r(t)$, $r'(t) \equiv \frac{dr}{dt}$, $p \equiv p(t)$, $e \equiv e(t)$, $\chi \equiv \chi(t)$ and $w \equiv w(t)$
are functions of the Kerr-Schild time coordinate $t$ and where $\chi(t)$ monotonically
increases with $t$. For geodesic motion, $p(t)$, $e(t)$, and $w(t)$ 
remain constant; they do, however, evolve under the influence of the self-force. 

\begin{figure}[t]
\begin{center}
\includegraphics[clip,width=8.5cm,angle=0]{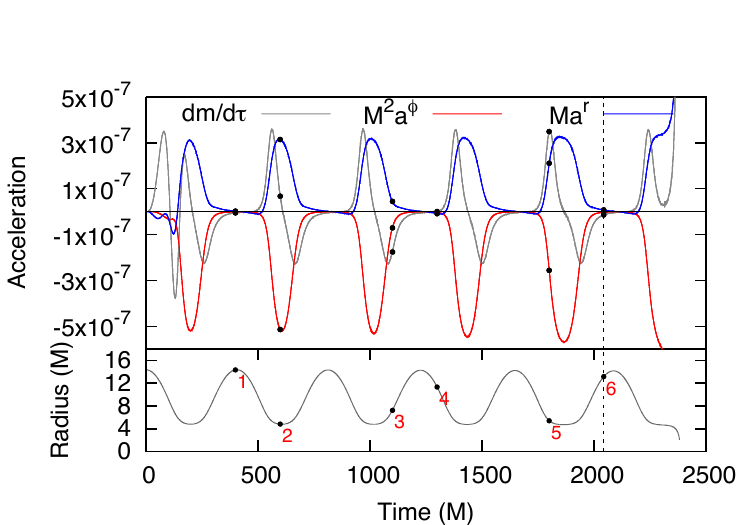}
\caption{(Color online) Acceleration along a trajectory that starts at $p=7.2, e=0.5$ for $q=1/32$.
The blue curve indicates $Ma^r$, the red curve $M^2a^\phi$, and the gray curve $\frac{dm}{d\tau}$.
The bottom plot shows the corresponding radial position of the particle.}
\label{fig:sf}
\end{center}
\end{figure}

The orbit we consider here starts at
$p_0 = 7.2, e_0 = 0.5$. In Fig.~\ref{fig:orbit}, we display an evolved orbit with
dimensionless charge $q := \bar{q}/M=1/32$, alongside a test-particle orbit for reference. Certain reference points along the 
orbit are identified to ease comparison with our other plots.
As initial data, we choose $\PhiR(t=0) =0$
and $\dot{\Phi}^{\rm R}(t=0) = 0$ and set the particle 
initially moving along the geodesic specified by $p_0 = 7.2, e_0 = 0.5$. This choice results 
in a burst of junk radiation, which contaminates the computed self-force at early times but 
eventually goes off the grid. By around $t=200M$, the 
evolved field $\PhiR$ settles down to give the 
appropriate geodesic self-force, as seen in Fig. \ref{fig:sf}. At $t=400M$, when the particle is very 
close to apoapsis, the computed self-force is allowed 
to act on the particle, and the system of equations (\ref{eqn:fieldEqn}), (\ref{eqn:EOM}), and 
(\ref{eqn:massEqn}) is evolved simultaneously for all subsequent times.
For this particular case ($q =1/32$), the particle makes approximately 
16 revolutions ($\sim4$ full radial cycles) 
before reaching the horizon. 

\begin{figure}[t!]
\begin{center}
\includegraphics[clip,width=8.5cm,angle=0]{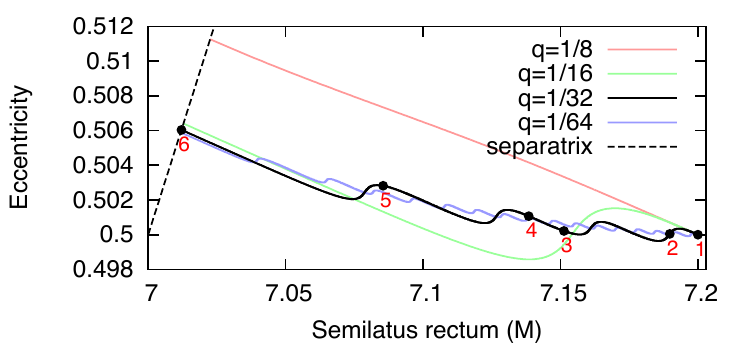}
\caption{(Color online) Orbital evolution in $p$--$e$ space, for an orbit that begins with $p=7.2,e=0.5$. Each oscillation 
in these tracks corresponds to one full radial cycle.}
\label{fig:pe5}
\end{center}
\end{figure}

Self-forced orbits can also be tracked in $p$--$e$ space. In 
Fig.~\ref{fig:pe5}, we observe that an
oscillating and secularly increasing eccentricity accompanies
the monotonic decrease in $p$. This secular increase is a generic feature of strong-field orbits 
under the influence of radiation reaction \cite{cutler-etal:94}.
The eccentricity oscillations, on the other hand, are a new 
feature of self-forced orbits not seen by flux-averaged 
models. They are due to the intrinsic periodicity in the 
local self-force that goes with the (quasi)periodic motion of the particle 
around the black hole. Indeed, it is easy to determine that 
$\dot{\tilde{E}} = -a^{\rm SF}_t$ and 
$\dot{\tilde{L}} = a^{\rm SF}_\phi$. A decrease in
$\tilde{E}$, while keeping $\tilde{L}$ constant, 
leads to a decrease in $e$, while a decrease in $\tilde{L}$, keeping
$\tilde{E}$ constant, tends to increase $e$. The self-force always 
decreases both $\tilde{E}$ and $|\tilde{L}|$, but it
does so at different rates depending on where the particle is. 
This competition between periodically varying loss rates is 
what leads to the oscillatory behavior in $e$. 

Since $\dot{\tilde{E}}$ and $\dot{\tilde{L}}$ scale as $q^2$, we expect that,
starting from the same initial conditions, the time it takes a particle to reach 
the separatrix (equivalently, the number of radial cycles)
should scale approximately as $1/q^2$. This is confirmed by our results. 
For $q=1/8$, the particle crosses the separatrix and 
plunges before completing one radial cycle. For $q =1/16, 1/32$, and $1/64$,
the particle plunges after one, four, and 16 full radial cycles,
respectively.

\begin{figure}[t!]
\begin{center}
\includegraphics[clip,width=8.5cm,angle=0]{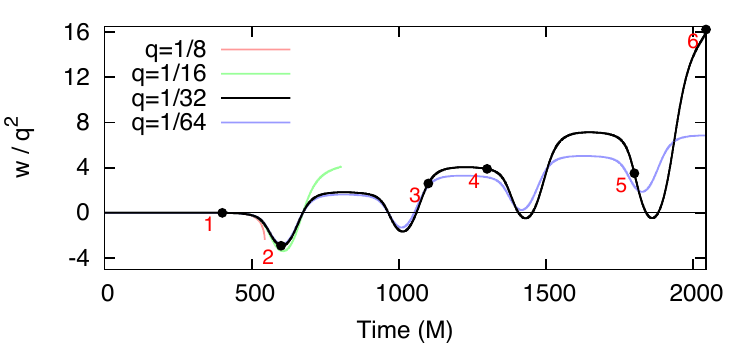}
\caption{(Color online) Positional element $w$ interpreted as a measure of the periapsis shift relative to its geodesic value. The curves end when the particle crosses the separatrix.}
\label{fig:w}
\end{center}
\end{figure}
It is difficult to disentangle dissipative and conservative 
effects of the self-force in a self-consistent evolution, since 
a reference geodesic \cite{mino:03,hinderer-flanagan:08,barack-sago:10}
or an explicit expression for the force \cite{pound-poisson:08b} 
is not available. Nevertheless, it is clear that the $p$--$e$ 
tracks do not fully characterize the orbital evolution. The osculating 
elements $\{p,e\}$ are in one-to-one correspondence with $\{\tilde{E},\tilde{L}\}$, whose 
rates of change are determined only by $a_t^{\rm SF}$ and 
$a_\phi^{\rm SF}$. The $r$ component of the self-acceleration cannot be 
inferred from the $p$--$e$ tracks alone, and instead its effect manifests 
in the secular change of the positional elements \cite{pound-poisson:08b}, 
an example of which is $w$ in Fig.~\ref{fig:w}. 
The observed changes in $w$ are completely missed by 
flux-averaged approximations.

The self-acceleration and mass change along the 
orbit is shown in Fig. \ref{fig:sf}. Only $a^r$ and $a^\phi$ are plotted here; the
third component $a^t$ is easily determined from the 
orthogonality condition $u_\alpha a^\alpha = 0$.
Inconsistent initial data contaminate the self-force early on, but this 
radiates away before self-consistent evolution starts at 
$t=400M$. The self-force depends most 
sensitively on the radial position of the particle, with its strength 
increasing the closer the particle is to the black hole. 
There is a small noticeable change in the extrema of the
self-force, but it is possible that this is mainly due to the 
small corresponding shifts in the extremal radii of the orbit.
These properties likely describe the geodesic
self-force as well; it will be instructive to compare
self-consistent and geodesic self-forces. 

Earlier we reported an upper bound of $2\%$ on the error in our
self-force calculation.
While this may appear sizable, we reiterate that this is merely an upper bound which 
is reached \emph{only} for very brief portions of the orbit.
Our agreement with the results of Ref.~\cite{haas:07} is, in fact, 
significantly better throughout most of the orbit. 
By performing a higher resolution run,
we found that the maximum amplitude of the error was halved. The corresponding
phase at separatrix crossing, however, changed only very slightly
($\approx 0.01\%$). This suggests that the noise in the self-force does not
significantly affect the phase evolution for the length of the runs we consider
here.

\emph{Discussion}:-- We summarize by emphasizing a few points. 
First, our time domain 3D code is versatile: It is not limited 
to low eccentricities, equatorial orbits, or even the Schwarzschild 
spacetime. An ongoing challenge is to devise more efficient ways to evaluate 
the complicated expression for the effective source in the Kerr spacetime. This would, for example, allow one
to check if the recently discovered Flanagan-Hinderer
resonances \cite{flanagan-hinderer:10} persist in a self-consistent orbit.
Second, our approach makes it possible to assess the adiabatic argument on which 
Ref.~\cite{warburton-etal:12} is based or, for that matter, any other proposal 
for the computation of self-forced orbits. Third, our code readily gives self-forced 
waveforms at ${\scri}^{+}$ (see Fig. \ref{fig:fluxes}). These waveforms did not require any post-processing 
after the computation of the orbit; instead, both the orbit and waveform are calculated simultaneously. 
Finally, it is possible to generalize our approach to the more important 
gravitational case. But in that context, we stress that care is needed in handling 
delicate gauge conditions \cite{gralla-wald:08} and possible instabilities that may be 
brought forth by the nonradiative low multipoles of the metric perturbation \cite{dolan:11}. This represents the 
next major phase of development for the effective source program.

As is to be expected from a 3D code, the computed self-force along an orbit is limited in accuracy compared to other methods. 
Further development will be 
required to improve on this with limited computational resources. 
Moreover, the code is too slow for the task of mass-producing waveforms that will 
sufficiently sample the entire EMRI parameter space. However, we emphasize that 
this is not the objective of our approach. The true value of our 
work lies in its ability to validate assumptions and predictions 
arising from all other (presumably faster) approximate methods. Our results provide the
first opportunity for these proposals to demonstrate that they indeed capture 
all the relevant features of self-consistent orbits and waveforms.

\textit{Acknowledgements:} 
The authors thank Eric Poisson, Abraham Harte,
Leor Barack, Sam Gralla, Luis Lehner, and 
Frank L\"offler for helpful comments and many fruitful discussions 
that helped shape this work and Roland Haas for sharing numerical data 
for the scalar self-force. Portions of this research were conducted with 
high performance computational resources provided by the Louisiana 
Optical Network
Initiative (http://www.loni.org/) and also used the Extreme Science and
Engineering Discovery Environment, which is supported by National
Science Foundation Grant No. OCI-1053575 (allocation TG-MCA02N014).
Some computations were also performed on the Datura
cluster at the Albert Einstein Institute.


%

\end{document}